\documentstyle[12pt]{article}
\input epsf
\begin{document}
%%%%%%%%%%%%%%%%%%%%%%%%%%%%%%%%% definitions %%%%%%%%%%%%%%%%%%%%%%%%%%%%%%%
\def\hw {\hbar \omega}
\def\mbs{\mbox{\boldmath$\sigma$}}
\def\gsim{\:\raisebox{-0.5ex}{$\stackrel{\textstyle>}{\sim}$}\:}
\def\lsim{\:\raisebox{-0.5ex}{$\stackrel{\textstyle<}{\sim}$}\:}
\def\pdM{\frac{\partial \; }{\partial M^2_{\pm} }}
\def\pdrnm{\frac{\partial \; }{\partial r_{nm}} }
\def\Ket#1{||#1 \rangle}
\def\Bra#1{\langle #1||}
\def\ubp{\bar{u}_{{\rm p}}}
\def\vbp{\bar{v}_{{\rm p}}}
\def\ubn{\bar{u}_{{\rm n}}}
\def\vbn{\bar{v}_{{\rm n}}}
\def\up{u_{{\rm p}}}
\def\vp{v_{{\rm p}}}
\def\un{u_{{\rm n}}}
\def\vn{v_{{\rm n}}}
\def\rop{\rho_{{\rm p}}}
\def\ron{\rho_{{\rm n}}}
\def\mbn{\mbox{\boldmath$\nabla$}}
\def\sss{\scriptscriptstyle}
\def\gp{g_{\sss +}}
\def\gm{g_{\sss -}}
\def\endauthors{}
\def\authors#1\endauthors{#1}
%%%%%%%%%%%%%%%%%%%%%%%%%%%%%%%%%%%%%%%%%%%%%%%%%%%%%%%%%%%%%%%%%%%%%%%%%%%%
%%%%%%%%%%%%%%%%%%%%%%%%%%%%%%%%%%%%%%%%%%%%%%%%%%%%%%%%%%%%%%%%%%%%%%%%%%%%
\begin{titlepage}
\pagestyle{empty}
\baselineskip=21pt
\rightline{McGill/95-58}
\rightline{hep-ph/9511437}
\vskip .2in
\begin{center}
{\large{\bf Charged majoron emission in neutrinoless double beta decay}}
$^*$
\end{center}

\vskip .1in

\authors
\centerline{C.~Barbero${}^{ a\dagger}$, J.~M.~Cline${}^b$,
F.~Krmpoti\'{c}$^{a\dagger}$ and D.~Tadi\'c${}^c$}
\vskip .15in
\centerline{\it ${}^a$ Departamento de F\'\i sica, Facultad de Ciencias
Exactas}
\centerline{\it
Universidad Nacional de La Plata, C. C. 67, 1900 La Plata, Argentina.}
\vskip .1in
\centerline{\it ${}^b$ McGill University, Montr\'eal, Qu\'ebec H3A 2T8,
Canada.}
\vskip .15in
\centerline{\it ${}^c$ Physics Department, University of Zagreb}
\centerline{\it Bijeni\v cka c. 32-P.O.B. 162, 41000 Zagreb, Croatia.}
\endauthors
\vskip 0.5in
\centerline{ {\bf Abstract} }
\baselineskip=18pt
We examine in detail the predictions of the charged majoron model,
introduced recently by Burgess and Cline, for $0^+\rightarrow 0^+$
$\beta\beta$ transitions.  The relevant nuclear matrix elements are
evaluated, within the quasiparticle random phase approximation,
for $^{76}$Ge,~$^{82}$Se,~$^{100}$Mo,~$^{128}$Te and $^{150}$Nd
nuclei. The calculated transition rates turn out to be much
smaller than the experimental upper limits on possible majoron
emission, except in a small region of the model's parameter space.
\bigskip

\vspace{0.5in}
\noindent
$^*$Work partially supported by the Fundaci\'on Antorchas, Argentina,
the CONICET from Argentina, the Universidad Nacional de La Plata, NSERC
of Canada and FCAR of Qu\'ebec.\\ $^{\dagger}$Fellow of the CONICET
from Argentina.
\end{titlepage}
\baselineskip=18pt

In 1987 Elliott, Hahn and Moe \cite{Ell87} observed, for the
first time, the two-neutrino double beta decay ($\beta\beta_{2\nu}$) of
$^{82}$Se into $^{82}$Kr by a direct counting method. Almost
simultaneously, Vogel and Zirnbauer \cite{Vog86} showed that, within
the quasiparticle random phase approximation (QRPA), it was possible to
explain the smallness of the measured $\beta\beta_{2\nu}$ decay rates.
Since then, impressive progress has been achieved in the experimental
investigation, and the $2\nu$-decay mode has been unambiguously
observed in several nuclear isotopes \cite{Moe94}.  This process occurs
at second-order in the charged-current weak interactions in the
Standard Model, and is the slowest process measured so far in nature.
As such, it offers a unique opportunity to test the nuclear structure
models for half-lives $\gsim~ 10^{20}$ years.

However, the renaissance of interest in $\beta\beta$-decay over the
last decade is mostly stimulated by the possibility of observing other
decays to which these experiments are also sensitive. The hope is to
find a smoking gun for ``new physics" from beyond the standard model.
The most promising  processes of this type are the lepton-number
violating neutrinoless decay ($\beta\beta_{0\nu}$), and the decay
$\beta\beta_{M}$, in which the two outgoing electrons are accompanied
by a Nambu-Golstone boson, called the majoron. Both processes were
predicted \cite{Geo81,Doi88} by the model introduced by Gelmini and
Roncadelli \cite{Gel81}.  While this simple and elegant model
stimulated many experimental searches, it was subsequently found to be
incompatible with the LEP measurement of the invisible width of the $Z$
boson \cite{Abr89,Gon89}.

Thus, to account for an anomalous excess of $\beta\beta$ events for
which some experimental groups had preliminary evidence \cite{Ell91},
Burgess and Cline recently advocated a new class of ``charged majoron''
models \cite{Bur93,Bur94} so-named because their majoron carries the
$U(1)$ charge of lepton number, presumed to be unbroken.
Refs.~\cite{Bur93,Bur94} estimated the nuclear matrix elements needed
for charged majoron emission to account for the anomalous events
observed in $^{76}$Ge,~$^{82}$Se,~$^{100}$Mo and $^{150}$Nd nuclei. In
the present work, we compute the transition rate in detail for the
particular model given in \cite{Bur94}, simplifying their analytic
expressions, and evaluating the corresponding nuclear matrix elements
for the above-named elements as well as for $^{128}$Te.  We use the
quasiparticle random phase approximation (QRPA) \cite{Hir90,Hir90a},
which has been shown to give good estimates for the $\beta\beta_{2\nu}$
transition probabilities.  Finally, the resulting transition
probabilities are compared with the present experimental data.

In the charged majoron model (CMM) of ref.~\cite{Bur94}, the standard
model gauge symmetry group  is augmented by a global nonabelian flavour
symmetry group, $SU_{ F}(2) \times U_{ L'}(1)$, which breaks down to
the ordinary lepton number $U_{L}(1)$ subgroup (see also refs.
\cite{Car93,Bam95}). To implement this symmetry-breaking pattern, an
electroweak-singlet, $SU_F(2)$-doublet scalar field $\Phi$ is
introduced, which gets a vacuum expectation value.  The model also
includes nonstandard electroweak-singlet Majorana neutrinos $(N_{\sss
+},N_{\sss -})$, $s_{\sss +}$ and $s_{\sss -}$, and the resulting
lagrangian density which respects all the symmetries is
\begin{equation}
\label{1}
{\cal L} = - \lambda \bar{L}\tilde{H}P_R s_{\sss -}
- M \bar s_{\sss +} P_R  s_{\sss -} -  \gp \, (\bar{N} P_L  s_{\sss +}) \;
\Phi - \gm \, (\bar{N} P_L  s_{\sss -}) \; \tilde{\Phi} + h.c. .
\end{equation}
Here $P_L$ and $P_R$ denote the projections onto left- and right-handed
spinors, $L$ is the usual lepton doublet, and $\tilde{H}$ is the charge
conjugate of the Higgs doublet.  After symmetry breaking by the vacuum
expectation values $\langle H\rangle = v = 174$ GeV, $\langle
\Phi\rangle = u\sim 100$ MeV, the resulting neutrino mass matrix yields
a massless neutrino $\nu'_e$ and two heavy Dirac neutrinos $\psi_{\pm}$
with masses
\begin{equation}
\label{2}
M_{\sss \pm}^2=\widetilde{M}^2\pm\sqrt{\widetilde{M}^4-\gp^2u^2
(\lambda^2v^2+\gm^2u^2)};~~
\widetilde{M}^2=\frac{1}{2}\left[M^2+\lambda^2v^2+(\gp^2+\gm^2)u^2\right].
\end{equation}
In terms of the mass eigenstates, the electroweak eigenstate is
\begin{equation}
	|\nu_e\rangle = c_\theta|\nu'_e\rangle + s_\theta c_{\alpha} 
	|\psi_{\sss -}\rangle + s_\theta s_{\alpha}|\psi_{\sss +}\rangle
\label{3}\end{equation}
where $s_{\theta}=\sin\theta$, $c_{\theta}=\cos\theta$, {\it etc.,} denote the
mixing angles with
\begin{equation}
\label{4}
\tan\theta=\frac{\lambda v}{\gm u};~~~
\tan2\alpha=\frac{2M\sqrt{\lambda^2v^2+\gm^2u^2}}
{M^2-\lambda^2v^2+(\gp^2-\gm^2)u^2}.
\end{equation}

The transition probability for the $0^+\rightarrow 0^+$ $\beta\beta$-decay
has the form \cite{Bur94}
\begin{equation}
\label{5}
\Gamma(\beta\beta)
=\frac{(G_F \cos\theta_C)^4}{256\pi^5}\left|{\cal A}(\beta\beta)\right|^2
\int(Q-\epsilon_1-\epsilon_2)^3\prod_{i=1}^2 k_i\epsilon_i
F(\epsilon_i)d\epsilon_i,
\end{equation}
in the notation of ref.~\cite{Bur94}, with
the transition amplitude given by
\begin{equation}
\label{6}
{\cal A}(\beta\beta)=2\sqrt{2} (s_{2\alpha} s_{2\theta} s_\theta)i\gp g^2_A
\Bra{0^+}
	\sum_{nm} {\bf Y}_{Rnm}\cdot{\hat{\bf r}}_{nm}\,
	\partial h_{\alpha}/\partial r_{nm}\Ket{0^+}.
\end{equation}
Here
\begin{eqnarray}
\label{7}
h_{\alpha}(r_{nm})&=& i\mu
\int\frac{d^4k}{(2\pi)^4}\frac{e^{-i{\bf k}\cdot{\bf r}_{nm}}} 
{(k_0^2-\mu^2+i\epsilon)}\,(P_{\sss -} - P_{\sss +})(P_0 - s_\alpha^2P_{\sss +} 
- c_\alpha^2 P_{\sss -});
\nonumber\\
P_i &=& (k^2-M_i^2+i\epsilon)^{-1};\quad M_0\equiv 0,
\end{eqnarray}
with ${\bf r}_{mn}$ and $\mu$ being, respectively, the separation in
the position between two decaying nucleons and the average excitation
energy of the intermediate nuclear states, and
\begin{equation}
\label{8}
{\bf Y}_{Rnm}=i\left[(\mbs_nC_m-C_n\mbs_m)
+i\left[\frac{g_{\sss V}}{g_{\sss A}}\right](\mbs_n\times{\bf D}_m+
{\bf D}_n\times\mbs_m)
+\left[\frac{g_{\sss V}}{g_{\sss A}}\right]^2({\bf D}_n-{\bf D}_m)\right],
\end{equation}
%is the same nucleon recoil operator\footnote{except for the sign of the
%second term, which we believe to be due to an error in
%ref.~\cite{Doi88}} as appears in Doi {\it et al.,} \cite{Doi88}.
happens to be same nucleon recoil operator as appears in Doi {\it et
al.,} \cite{Doi88}, except for the sign of the second term.  (We
believe the sign difference is due to an error in ref.~\cite{Doi88}.)

It is possible to simplify the above expressions for the matrix
element.  Previously it has been argued that the middle term of
eq.~(\ref{8}) is the most important. Within the approximation where
the momenta of electrons and majoron are neglected in comparison with those
of the nucleon, and because the main
contribution involves the spin singlet state of two nucleons,
this term gives \cite{Doi88,Doi85}
\begin{equation}
\label{9}
{\bf Y}_{Rnm}\cdot \hat{\bf r}_{nm}\simeq \frac{f_R}{2M_N}(\mbs_n\cdot\mbs_m)
(\mbn_{\!nm}\cdot \hat{\bf r}_{nm}),
\end{equation}
where $M_N$ is the nucleon mass, $f_R\simeq 5.6$ (for $g_A=1$ \cite{Hir90,
Bro85})
and $\mbn_{\!nm} = \partial/\partial{\bf r}_{nm}$.  From eq.~(\ref{6}) it is
clear that the amplitude is largest if $s_{2\alpha}=1$, so we shall
make that assumption.  Furthermore the mixing angle $\theta$ is
typically constrained to be small, so we take $s_\theta = \theta$.
Then, after performing the $p_0$ integral in eq.~(\ref{7}) and using the
identity $\mbn\cdot \hat{\bf r}\partial h_{\alpha}(r)
/\partial{r} = \mbn^2 h_{\alpha}(r) $, the amplitude can be
written as a difference of two pieces,
\begin{equation}
\label{10}
   {\cal A}(\beta\beta)=\frac{i\theta^2 g^2_A \gp}{\sqrt{2}\pi}
	{\cal M}_{\sss CM};~~~
	{\cal M}_{\sss CM}=     {\cal M}^+_{\sss CM} - {\cal M}^-_{\sss CM},
\end{equation}
where
\begin{equation}
	{\cal M}^{\pm}_{\sss CM} = \frac{f_R}{2M_N}
	\Bra{0^+}\sum_{mn}\tilde{h}(r_{mn};M_{\pm})\,
	\mbs_n\cdot\mbs_m\Ket{0^+},
\label{11} \end{equation}
is the nuclear matrix element for charged majoron emission 
corresponding to the exchange of the neutrino with mass $M_{\sss \pm}$, and
\begin{equation}
	\tilde{h}(r_{mn};M_{\sss \pm})=\frac{1}{M_{\sss \pm}^2}
	\left[h(r_{mn};M_{\sss \pm})-h(r_{mn};0) \right]
	+\frac12\pdM h(r_{mn};M_{\sss \pm})
\label{12} \end{equation}
is the corresponding neutrino potential, with
\begin{equation}
h(r_{mn};M_{\sss \pm})=\int\frac{d{\bf k}}{2\pi^2}
e^{-i{\bf k}\cdot{\bf r}_{mn}}
\frac{k^2}{\omega_{\sss \pm}(\omega_{\sss \pm}+\mu)};~~~
\omega_{\sss \pm}=\left(k^2+M_{\sss \pm}^2\right)^{1/2}.
\label{13} \end{equation}

The inverse half-life can be now cast in the form
\begin{equation}
[T(0^+\rightarrow 0^+)]^{-1}=g_{\sss CM}^2 {\cal G}_{\sss CM}
	|{\cal M}_{\sss CM}|^2,
\label{14} \end{equation}
where
\begin{equation}
g_{\sss CM}= \gp\theta^2/2, 
\label{15} \end{equation}
is the effective majoron coupling, and
\begin{equation}
{\cal G}_{\sss CM}= \frac{(G_Fg_{\sss A}\cos\theta_C)^4}{128\pi^7\ln 2}
\int(Q-\epsilon_1-\epsilon_2)^3\prod_{i=1}^2 k_i\epsilon_i F(\epsilon_i)
d\epsilon_i,
\label{16} \end{equation}
is the kinematical factor as defined in ref. \cite{Doi88}.

To perform the nuclear structure calculation we use the Fourier-Bessel
expansion of the charged majoron matrix element. Thus
\[
{\cal M}_{\sss CM}^{\pm} = \sum_{LSJ^{\pi} }m(M_{\sss \pm};L,S,J^{\pi}),
\]
where $L$, $S$, $J$ and $\pi$ are, respectively, the orbital angular
momentum, the spin, the total angular momentum and the parity of the
intermediate nuclear states.  Within the QRPA formulation presented in
ref. \cite{Hir90} the individual nuclear matrix element is
\begin{equation}
m(M_{\sss \pm};L,S,J^{\pi}) =(-)^{S} \sum_{\alpha {\rm pnp}'{\rm n}'}
 W^{LSJ}_{{\rm pn}}  W^{LSJ}_{{\rm p}'{\rm n}'}
{\cal R}^{L} ({\rm pn}{\rm p}'{\rm n}';M_{\sss \pm})
\Lambda_{+}({\rm pn};\alpha J^{\pi})
\Lambda^{*}_{-}({\rm p}'{\rm n}';\alpha J^{\pi}).
\label{17} \end{equation}
The amplitudes $\Lambda_{\sss \pm} ({\rm pn};\alpha J^{\pi})$  are
\begin{eqnarray*}
\Lambda_{+}({\rm pn};\alpha J)=\sqrt{\rop \ron}
\left[\up \vn X_{{\rm pn};\alpha J}+\vbp \ubn Y_{{\rm pn};\alpha
J}\right];
\\
\Lambda_{-}({\rm pn};\alpha J)=\sqrt{\rop \ron}
\left[\vbp \ubn X_{{\rm pn};\alpha J}+\up \vn Y_{{\rm pn};\alpha J}\right],
\end{eqnarray*}
where the unbarred (barred) quantities indicate that the quasiparticles
are defined with respect to the initial (final) nucleus;
$\rop^{-1}=\up^{2}+\vbp^{2}$, $\ron^{-1}=\ubn^{2}+\vn^{2}$,
and all the remaining notation has the standard meaning \cite{Hir90}.
The angular momentum and radial pieces in (\ref{17}) are, respectively,
\[
W^{LSJ}_{{\rm pn}} = i^{\ell_{{\rm n}}-\ell_{{\rm p}}+L}\sqrt{2}\, \hat{J}\, 
\hat{S}
\,\hat{L} \,\hat{j}_{{\rm p}} \,\hat{j}_{{\rm n}}\, \hat{\ell}_{{\rm n}}\,
(\ell_{{\rm n}} 0 L 0\mid \ell_{{\rm p}} 0)
\left\{ \begin{array}{ccc}
\ell_{{\rm n}} & \frac{1}{2} & j_{{\rm n}} \\
L        &    S        &  J    \\
\ell_{{\rm p}} & \frac{1}{2} & j_{{\rm p}} \\
\end {array} \right\} ,
\]
and
\[
{\cal R}^{L}({\rm pn}{\rm p}'{\rm n}';M_{\sss \pm})=
\int^{\infty}_{0} dkk^{2} {\rm v}(k;M_{\sss \pm})
{\sf R}^L_{{\rm pn}}(k){\sf R}^L_{{\rm p}'{\rm n}'} (k),
\]
with
\[
{\rm v}(k;M_{\sss \pm})
=\frac{2k^2}{\pi}\left\{\frac{1}{M_{\sss \pm}^2}\left[
\frac{1}{\omega_{\sss \pm}(\omega_{\sss \pm}+\mu)}-
\frac{1}{k(k+\mu)}\right] + \frac12\pdM\frac{1}{\omega_{\sss \pm}
(\omega_{\sss \pm}+\mu)}
\right\}
\]
and
\[
{\sf R}^L_{{\rm pn}}(k)=\int^{\infty}_{0}u_{\rm n}(r) 
u_{\rm p}(r)j_L(kr)r^{2}dr,
\]
{\it u}(r) being the single-particle radial wave functions.

The numerical calculations were performed with the $\delta$-force (in
units of MeV fm$^{3}$) $ V=-4\pi({\it v}_{s}P_{s}+{\it
v}_{t}P_{t})\delta(r), $ with different strength constants ${\it
v}_{s}$ and  ${\it v}_{t}$ for the particle-hole, particle-particle and
pairing channels.  An eleven-dimensional model space was used,
including all the single particle orbitals of oscillator shells $3\hw$
and $4\hw$, plus the $0h_{9/2}$ and $0h_{11/2}$ orbitals from the
$5\hw$ oscillator shell. The single particle energies, as well as the
parameters ${\it v}_{s}^{pair}({\rm p})$ and  ${\it v}_{s}^{pair}({\rm
n})$, have been fixed by the procedure employed in ref.~\cite{Hir90}
({\it i.e.,} by fitting the experimental pairing gaps  to a Wood-Saxon
potential well).

The dependence of ${\cal M}_{\sss CM}^{\sss \pm}$ on $M_{\sss \pm}$ is
illustrated in Fig.\ 1 for the $^{76}$Ge$\rightarrow ^{76}$Se decay.
As one might expect, the matrix element is insensitive to the mass
until it starts to exceed the Fermi momentum of the nucleons, around
100 MeV, thereafter giving a $1/M^2$ suppression.  Since the matrix
element ${\cal M}_{\sss CM}$ is the difference between the $M_{\sss +}$
and $M_{\sss -}$ contributions, one can easily recover the result for
an arbitrary pair of masses by taking the difference between the two
corresponding matrix elements.  Obviously for $M_{\sss -}\cong M_{\sss
+}$ there is destructive interference between the matrix elements
${\cal M}^{\sss -}_{\sss CM}$ and ${\cal M}^{\sss +}_{\sss CM}$.  This
happens, for example, with the choice of parameters $M=\lambda v$ and
$g_{\sss \pm}\sim 1$, which from eqs.\ (\ref{2}) and (\ref{4}) yields
$M_{\sss \pm}=g_{\sss +}u\sqrt{1\pm\theta}$.  Since the mixing angle
$\theta$ is experimentally constrained to be of the order of
$0.1$,\footnote{For a discussion of these experimental constraints, see
ref.~\cite{Bur93}.} the calculated values of $[T_{1/2}^{CM}]^{-1}$ turn
out to be four to five orders of magnitude smaller than the
corresponding experimental upper limits for the majoron emission, due
to the additional $\theta^2$ suppression.  These limits are displayed
in Table 1 for the $^{76}$Ge, $^{82}$Se, $^{100}$Mo, $^{128}$Te and
$^{150}$Nd nuclei.  For the sake of completeness, the table also shows
the measurements for the $\beta\beta_{2\nu}$ decays, and ${\cal
G}_{\sss CM}$ for the effective axial-vector coupling $g_{\sss A}=1$
\cite{Hir90,Bro85}.

%%%%%%%%%%%%%%%%%%%%%%%%%%%%%%%%%%%%%%
%%%  figure 1
%%%%%%%%%%%%%%%%%%%%%%%%%%%%%%%%%%%%%%
\vskip 1truecm
\epsfbox{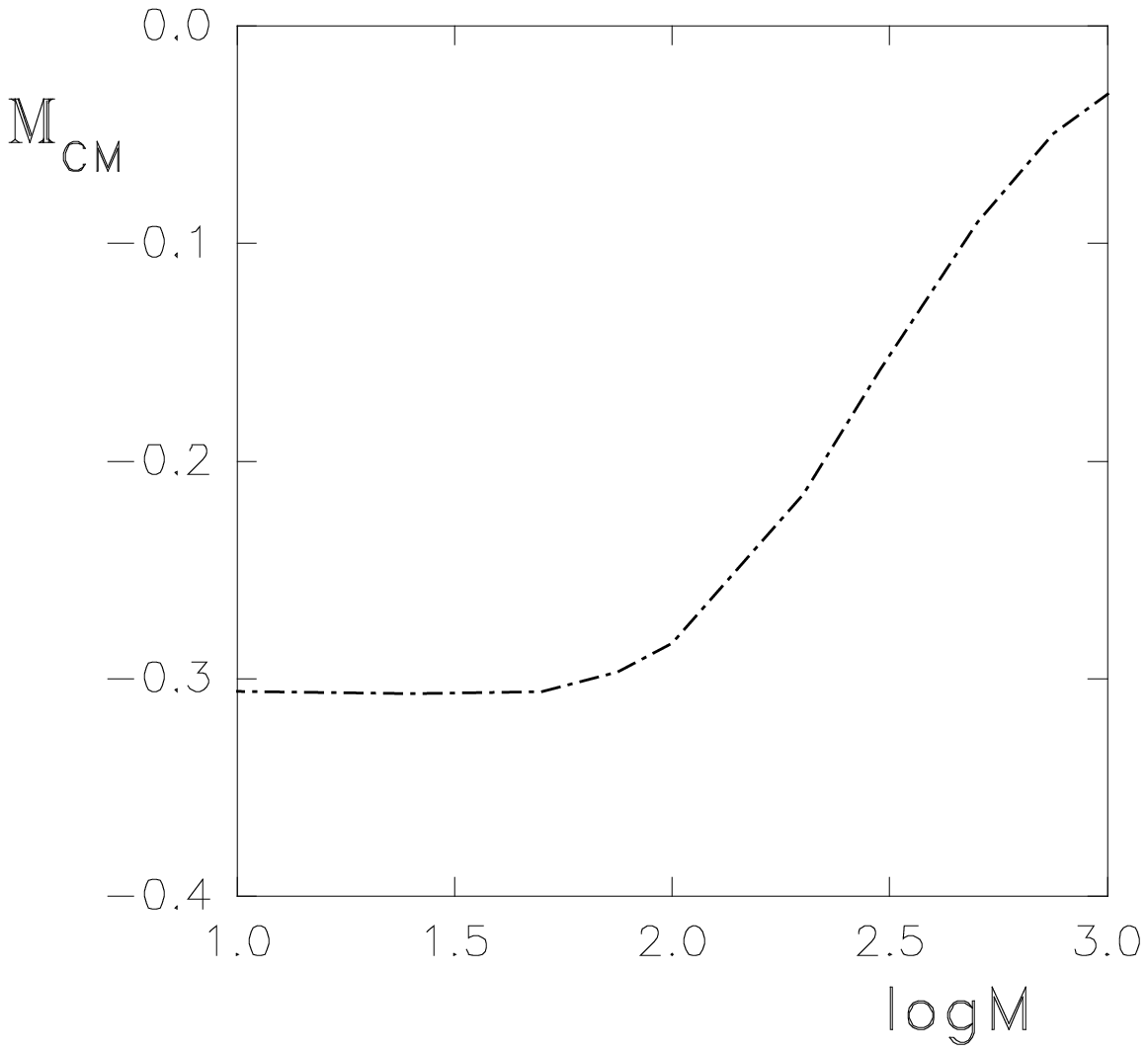}
\baselineskip 17pt\noindent
Figure 1: Charged majoron matrix element ${\cal M}_{\sss CM}^{\sss \pm}$ 
(in natural units)
for the $^{76}$Ge$\rightarrow ^{76}$Se decay,
as a function of the heavy neutrino mass $M_{\sss\pm}$ (in units of MeV).
\vskip0.5truecm
\baselineskip 20pt

Figure 1 also shows that the most favorable situation for majoron
emission occurs  for in limit that the heavier neutrino becomes
infinitely massive, thus making no contribution to the total rate.  For
$M_{\sss-}$ we will use $100$~MeV, since this is the largest value
which still gives an unsuppressed amplitude.  In Table 1 we compare the
theory to the data by showing how large the effective coupling $g_{\sss
CM}$ would have to be in this case in order for the CMM rate to be
equal to the experimental limit on the majoron-emitting mode of
$\beta\beta$ decay.  For all the elements, $g_{\sss CM}$ must be
$0.1-0.2$ to be presently observable.  Given the above mentioned
experimental constraints $\theta$, from eq.~(\ref{15}) it is clear that
one would need a strong coupling in the sterile neutrino sector in
order to achieve such a large value of the majoron-emitting rate.

\begin{table}[t]
\begin{center} %{\large Table1 }
\caption {The size of the effective coupling $g_{\sss CM}$
which would be needed for the rate of emission of charged majorons to
be equal to the experimental limit on this process; ${\cal G}_{\sss
CM}$ and ${\cal |M|}_{\sss CM}$ are the corresponding kinematical
factors and the nuclear matrix elements, respectively. For the sake of
comparison the pertinent experimental data for the two-neutrino
processes are also shown.}
\label{tab1}
\bigskip
\begin{tabular}{l|ccccc}
\hline
\hline
Nucleus&${[T_{1/2}^{2\nu}]^{-1}_{exp} (yr^{-1})}$
&${[T_{1/2}^{M}]^{-1}_{exp} (yr^{-1})}$
&${\!{\cal G}_{\sss CM}(yr^{-1})\!}$&${\cal |M|}_{\sss CM}$
& $g_{\sss CM}$ needed\\
\hline
{$^{76}Ge$}&$\!(6.99 \pm 0.08)\;10^{-22}\,^{a)}\!$
&$\!<6.0\;10^{-23}\,^{a)}\!$&$\!2.10\;10^{-20}\!$&$0.28$
&$\!<0.19\!$\\
{$^{82}Se$}&$\!(9.26_{-2.23}^{+0.51})\;10^{-21}\,^{b)}\!$
&$\!<6.2\;10^{-22}\,^{c)}\!$&$\!3.55\;10^{-19}\!$&0.28
&$\!<0.15\!$\\
{$^{100}Mo$}&$\!(8.69_{-2.27}^{+1.51})\;10^{-20}\,^{d)}\!$
&$\!<3.0\;10^{-21}\,^{e)}\!$&$\!7.33\;10^{-19}\!$&0.31
&$\!<0.21\!$\\
{$^{128}Te$}&$\!(1.30\pm 0.07)\;10^{-25}\,^{f)}\!$
&$\!<1.3\;10^{-25}\,^{f)}\!$&$\!5.24\;10^{-22}\!$&0.27
&$\!<0.059\!$\\
{$^{150}Nd$}&$\!(5.88_{-3.46}^{+3.11}\pm 1.21)\;10^{-20}\,^{g)}\!$
&$\!<1.9\;10^{-21}\,^{h)}\!$&$\!6.06\;10^{-18}\!$&0.13
&$\!<0.077\!$\\
\hline \hline \end{tabular} \end{center}
$^{a})$ (laboratory data) ref.\ \cite{Bec93}\\
$^{b})$ (laboratory data) ref.\ \cite{Ell92}\\
$^{c})$ (laboratory data) ref.\ \cite{Ell87}\\
$^{d})$ (laboratory data) ref.\ \cite{Eji91}\\
$^{e})$ (laboratory data) ref.\ \cite{Als88}\\
$^{f})$ (geochemical data) ref.\ \cite{Ber92}\\
$^{g})$ (laboratory data) ref.\ \cite{Art93}\\
$^{h})$ (laboratory data) ref.\ \cite{Moe94a}\\
\end{table}

In conclusion, we have found that the rate of majoron-emitting,
neutrinoless $\beta\beta$ decay in the charged majoron model is
unobservably small unless there is large mixing between exotic sterile
neutrinos of mass $\gsim 100$ MeV, and strong couplings among the
sterile neutrinos. In computing the relevant nuclear matrix elements,
we have not considered the effects of finite nucleon size or
short-range two-nucleon correlations, which would tend to reduce the
calculated matrix elements \cite{Hir90}.  On the other hand, the
arguments used to simplify the nuclear transition operator from
(\ref{8}) to (\ref{9}) are not rigorous, and it is also possible that
future variants of the CMM considered here might evade the suppression
of the rate by the mixing angle $\theta$.  Thus, while we believe that
ours is the most quantitative analysis of the CMM to date, if the
experimental situation should give serious indications of anomalous
$\beta\beta$ decay events in the future, it would become appropriate to
undertake a yet more careful evaluation of the model's predictions.

\bigskip

Note added: As we were finishing this work, Hirsch {\it et al.}\
\cite{Hir95} presented results including a somewhat less detailed
analysis of the CMM, in which they reached conclusions similar to ours.


\bigskip\bigskip

\begin{thebibliography}{99}
\bibitem{Ell87} S.R. Elliot, A.A. Hahn and M.K. Moe, Phys. Rev. Lett.
{\bf{59}} (1987) 1649.
\bibitem{Vog86} P. Vogel and M.R. Zirnbauer, Phys. Rev. Lett. {\bf 57} (1986) 
731
\bibitem{Moe94} M. Moe and P. Vogel, Ann. Rev. Nucl. Part. Sci. {\bf {44}}
(1994) 247.
\bibitem{Geo81}  H. Georgi, S.L. Glashow and S. Nussinov,
Nuc. Phys. {\bf B193}  (1981) 297.
\bibitem{Doi88} M. Doi, T. Kotani and E. Takasugi, Phys. Rev. {\bf {D37}},
(1988), 2575
\bibitem{Gel81}  G.B.  Gelmini  and  M.  Roncadelli,
Phys. Lett.  {\bf B99}  (1981) 411.
\bibitem{Abr89} G.S. Abrams et al., Phys. Rev. Lett.  {\bf 63}  (1989) 724.
\bibitem{Gon89} M.C. Gonzalez-Garcia and Y. Nir,
Phys. Lett.  {\bf B232} (1989) 383.
\bibitem{Ell91} S.R. Elliot, M.K. Moe, M.A. Nelson and M.A. Vient, J. Phys.
{\bf{G17}} (1991) S145; Nucl. Phys. {\bf {B}} (Proc. Suppl.) {\bf{31}}
(1993) 68.
\bibitem{Bur93} C.P. Burgess and J.M. Cline, Phys. Lett. {\bf {B298}}, (1993),
141.
\bibitem{Bur94} C.P. Burgess and J.M. Cline, Phys. Rev. {\bf {D49}}, (1994),
5925.
\bibitem{Hir90}
J. Hirsch, E. Bauer and F. Krmpoti\'{c}, Nucl. Phys. {\bf A516} (1990) 304.
\bibitem{Hir90a} J. Hirsch and F. Krmpoti\'{c}, Phys. Rev. {\bf C41} (1990) 
	792;
J. Hirsch and F. Krmpoti\'{c}, Phys.  Lett.  {\bf B246}  (1990)  5;
F. Krmpoti\'{c} and S. Shelly Sharma, Nucl. Phys. {\bf A572} (1994) 329.
\bibitem{Car93} C.D. Carone, Phys. Lett. {\bf {B308}}, (1993), 85.
\bibitem{Bam95} P. Bamert, C.P. Burgess and R.N. Mohapatra, Nucl. Phys.
{\bf {B449}}, (1995), 25.
%\bibitem{Tom85} T. Tomoda, A. Faessler, K. W. Schmid and F. Gr\H{u}mmer,
\bibitem{Doi85} M. Doi, T. Kotani and E. Takasugi, Prog. Theor. Phys.
Suppl. {\bf {83}}, (1985), 1.
\bibitem{Bro85} B.A. Brown and B.H. Wildenthal, Atom. Data and Nucl. Data 
Tables {\bf{33}}, (1985), 347.
\bibitem{Bec93} M. Beck {\it{et al.}}, Phys. Rev. Lett. {\bf{70}}, (1993), 
2853.
\bibitem{Ell92}  S.R. Elliot, A.A. Hahn, M.K. Moe, M.A. Nelson and M.A. Vient,
Phys. Rev. {\bf C46}, (1992), 1535.
\bibitem{Eji91} H. Ejiri, K. Fushimi, T. Kamada, H. Kinoshita, H. Kobiki,
 H. Ohsumi,
K. Okada, H. Sano, T. Shibata, T. Shima, N. Tanabe, J. Tanaka, T. Taniguchi,
T. Watanabe and N. Yamamoto, Phys. Lett. {\bf B258}, (1991), 17.
\bibitem{Als88} M. Alston-Garnjost {\it{et al.}}, Phys. Rev. Lett. {\bf{60}},
(1988), 1928.
\bibitem{Ber92} T. Bernatowicz, J. Brannon, R. Brazlle, R. Cowsik, 
C. Hohenberg and F. Podosek, Phys. Rev. Lett. {\bf 69}, (1992), 2341;
T. Bernatowicz {\it{et al.}}, Phys. Rev. {\bf C47}, (1993), 806.
\bibitem{Art93}  V. Artemiov {\it{et al.}}, JEPT Lett. {\bf 58}, (1993), 1009.
\bibitem{Moe94a} M.K. Moe, M.A. Nelson and M.A. Vient, Prog. Part. Nucl. Phys.
{\bf 32} (1994) 247.
\bibitem{Hir95}  M. Hirsch, H.V. Klapdor--Kleingrothaus, S.G. Kovalenko and H.
   P\"as, preprint hep-ph/9511227. 
\end{thebibliography}
\end{document}